\documentclass[prd, onecolumn, nofootinbib, floatfix, notitlepage,longbibliography]{revtex4-2}

\usepackage{amsmath}
\usepackage{graphicx}
\usepackage{dcolumn}
\usepackage{bm}
\usepackage{epsfig}
\usepackage{amssymb,latexsym,mathrsfs}
\usepackage{graphicx}
\usepackage{color}
\usepackage{xcolor} 
\usepackage{mathtools} 
\usepackage{hyperref}

\hypersetup{
    colorlinks=true,
    linkcolor=red,
    citecolor=blue,
} 

\newcommand{\be}{\begin{equation}}
\newcommand{\ee}{\end{equation}}
\newcommand{\bs}{\begin{split}} 
\newcommand{\bea}{\begin{eqnarray}}
\newcommand{\eea}{\end{eqnarray}}

\newcommand{\eps}{\epsilon} 
\newcommand{\Ok}{\Omega_k} 
\newcommand{\Om}{\Omega_m} 
\newcommand{\sigok}{\sigma(\Ok)}

\begin{document}

\title{Testing Cosmology with Double Source Lensing} 

\author{Divij Sharma${}^{1}$, Thomas E. Collett${}^{2}$, Eric V. Linder${}^{3,4}$} 
\affiliation{
${}^1$Department of Physics, 
University of California, Berkeley, CA 94720, USA\\
${}^2$Institute of Cosmology and Gravitation, University of Portsmouth, Portsmouth, PO1 3FX, UK\\ 
${}^3$Berkeley Center for Cosmological Physics \& Berkeley Lab, 
University of California, Berkeley, CA 94720, USA\\ 
${}^4$Energetic Cosmos Laboratory, Nazarbayev University, Astana, Qazaqstan 010000\\ 
} 

\begin{abstract}
Double source lensing provides a dimensionless ratio of 
distance ratios, a ``remote viewing'' of cosmology through 
distances relative to the gravitational lens, beyond the 
observer. We use this to test the cosmological framework, 
particularly with respect to spatial curvature and the 
distance duality relation. We derive a consistency equation 
for constant spatial curvature, allowing not only the 
investigation of flat vs curved but of the 
Friedmann-Lema{\^\i}tre-Robertson-Walker framework itself. 
For distance duality, we demonstrate that the evolution of 
the lens mass profile slope must be 
controlled to $\gtrsim5$ times tighter fractional precision 
than a claimed distance duality violation. 
Using {\sc LensPop} forecasts of double source lensing 
systems in Euclid and LSST surveys we also explore constraints 
on dark energy equation of state parameters and any 
evolution of the lens mass profile slope. 
\end{abstract} 

\date{\today} 

\maketitle

\section{Introduction} \label{sec:intro}

The Friedmann-Lema{\^\i}tre-Robertson-Walker (FLRW) framework 
for describing the cosmology of our universe is highly 
successful. Considerable effort is dedicated toward 
determining the parameters of the cosmology, e.g.\ the 
energy densities of the components and their equations 
of state \cite{2201.08666}. Another path for cosmological investigation is 
to test the framework itself, e.g.\ testing whether gravity 
follows general relativity, whether spatial sections have 
constant curvature (flat or otherwise), and whether 
photons propagate along null geodesics. 
This can include a wide	variety	of techniques such
as testing geometric quantities	vs growth quantities, e.g.
\cite{1804.04320,2201.07025,2212.05003}.

Here we focus on testing the broad framework, and in particular probing 
spatial curvature in a general manner as well as photon 
propagation in terms of the distance duality relation. 
A variety of probes can be used in such tests, with systematics needing to	be stringently controlled
for each probe.	The 
more independent of cosmological and astrophysical parameters a probe is, the 
more general its conclusions may be. 
We therefore focus primarily on model independent 
constraints rather than parameter 
estimation (see e.g.\ \cite{2008.11286}
for a review of	parametric estimates of	curvature), 
and geometric probes such as 
distance measures. 

One distance measure 
stands out by being dimensionless and hence 
not dependent on an absolute scale to 
``anchor'' it: 
double source lensing (DSL; \cite{collett2012}). With DSL one can form a 
dimensionless quantity $\beta$ from the 
observable image positions associated with each source strongly gravitationally lensed 
by a common foreground galaxy. Such a quantity is a ratio 
of distance ratios {\it seen by the lens\/}, i.e.\ a 
``remote viewing'' of the universe, and furthermore is 
independent of the Hubble constant and fairly insensitive 
to the exact lens mass model (and even lens and source 
redshifts to some extent). As a geometric probe, it is 
well suited to testing the spacetime framework. Our goal is to supplement, not 
supplant, other probes, and enable a crosscheck 
in a more model independent manner.  

In Section~\ref{sec:double} we briefly review double  
source lensing. Section~\ref{sec:forecast} forecasts expectations for the data set that upcoming wide-field surveys will deliver. Section~\ref{sec:DEconstraints} explores the cosmological parameter leverage of DSL in constraining dark energy using the DSL forecasts. In Section~\ref{sec:curv} we 
apply DSL to probing spatial curvature, deriving a consistency 
test able to decide between a flat universe, one with constant 
curvature, and one that breaks the FLRW framework. 
Section~\ref{sec:dual} turns to the question of photon 
propagation and the distance duality relation, in particular 
examining the systematics requirements for measuring a 
violation. We summarize and conclude in Section~\ref{sec:concl}.

\section{Double Source Lensing} \label{sec:double} 

Lensed image positions, related to the Einstein radii of the 
circle of light imaged from a source aligned with the 
lens and observer, are observables. They depend on 
the mass of the lens and cosmological distances. Therefore, if one knows the mass of a lens, the Einstein radius is a direct geometrical probe of cosmology.  However astrophysical lenses are galaxies or bigger and the masses of such objects are rarely known with sufficient precision to break the mass-cosmology degeneracy inherent to the Einstein radius. Where multiple sources are present at different distances behind the same lensing mass this degeneracy can be lifted by taking the 
ratio of light deflection angles (Einstein radii in special 
cases) 
\citep{collett2012}. At high precision, the method also requires understanding  the lens density profile and of perturbative lensing from the sources and other mass along the line of sight. For a two-source plane system with one primary lens, the lens equations for the sources are 
\bea 
\mathbf{y} &=& \mathbf{x} - \beta_{12} \bm{\alpha}_l (\mathbf{x})\\ 
\mathbf{z} &=& \mathbf{x} - \bm{\alpha}_l (\mathbf{x}) - \bm{\alpha}_{s1}(\mathbf{x} - \beta_{12} \bm{\alpha}_l (\mathbf{x}))\ , 
\eea 
where $\bf x$ are positions on the image plane (as seen by the observer), $\bf y$ ($\bf z$) is the unlensed position of the first (second) source, $\bm{\alpha}_l({\bf x})$ is the reduced deflection at $\bf x$ caused by the primary lens acting on the final source plane, and 
$\bm{\alpha}_{s1}$ is the lensing deflection effect of mass of the first source. The 
primary quantity of cosmological interest, 
the ratio of distance ratios $\beta$ will 
be used as the DSL probe, with 
\begin{equation}
\beta_{12}\equiv \beta(z_l,z_{s1},z_{s2})=
\frac{r_{ls}(z_l,z_{s1})}{r_s(z_{s1})}\frac{r_s(z_{s2})}{r_{ls}(z_l,z_{s2})} \ , \label{eq:betadef} 
\end{equation}
where $r$ are the comoving distances. 
Note that $\beta$ is the cosmological scaling factor relating the reduced deflection angles between two source planes (the physical deflection caused by the lens is independent of the source redshift). For a singular isothermal sphere lens mass profile this is  the ratio of Einstein radii, but $\beta$ is more fundamental than that as it is the term that enters into the multiplane lens equations above regardless of deflector mass profiles. 
It is measured during the fitting of 
a multiplane lens model to the imaging data. 
The reduced sensitivity of $\beta$ to 
the mass profile(s) is key, as that is the 
dominant systematic in lensing. 
And as outlined in Sec.~\ref{sec:intro} and 
expanded upon in Secs.~\ref{sec:curv} and 
\ref{sec:dual}, its dimensionless and 
``remote viewing'' nature are particularly 
valuable for model independent tests. 

\begin{figure}[h]
\centering 
\includegraphics[width=0.75\columnwidth]{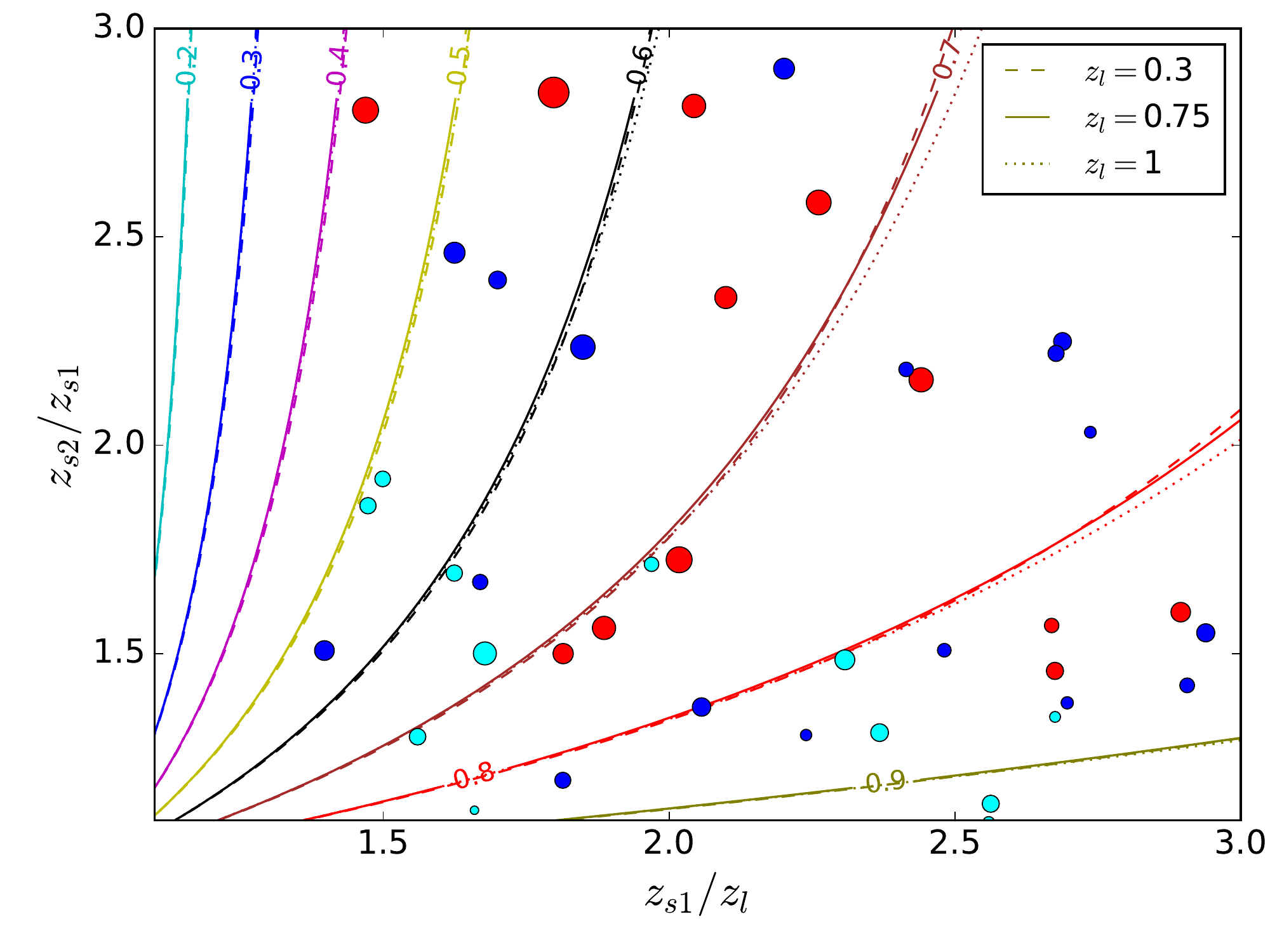}
\caption{Isocontours of $\beta$ are plotted 
vs $z_{s1}/z_l$ and $z_{s2}/z_{s1}$ 
for three lens redshifts $z_l=0.3$, 0.75, and 1.  
Note the contours vary very little with lens redshift $z_l$ for fixed ratios $z_{s1}/z_l$ and 
$z_{s2}/z_{s1}$.
The DSL parameter $\beta$ shows a 
broad flat valley between $0.6$--$0.8$ 
over a wide range of parameter space. 
The circles represent the mock data introduced in 
Secs.~\ref{sec:forecast} and \ref{sec:DEconstraints}, 
leaving off those outside the plot bounds. The radius of 
the circles is proportional to the simulated $1/\sigma(\beta)$, 
so more precise data are more prominent, and the colors 
correspond to $z_l=[0.2,0.3]$ (red), [0.3,0.5] (blue), 
and $z_l>0.5$ (cyan). 90\% of the data have 
$\sigma(\beta)<1.7\%$. 
}
\label{fig:isocontours} 
\end{figure}

Figure~\ref{fig:isocontours} shows how $\beta$ 
depends on the lensing system redshifts $z_l$, 
$z_{s1}$, $z_{s2}$. For a broad range of parameter 
space $\beta$ stays near 0.6--0.8. 
We also see that for given ratios $z_{s1}/z_l$ and 
$z_{s2}/z_{s1}$ the quantity $\beta$ is quite 
insensitive to the lens redshift. 
DSL also has good complementarity with time delay 
distances, also from strong lensing, as seen in 
\cite{1605.04910}, is a valuable crosscheck on 
the use of time delay lensing in testing the 
cosmological framework \cite{1802.04816}, and has 
interesting leverage at high redshift \cite{2204.03020}.

\section{Forecasting the future double source plane population}\label{sec:forecast}

Currently, only a small sample of galaxy scale DSL systems is known, and only the so-called Jackpot \citep{gavazzi2008,sonnenfeld2012,collett2020} has been used to precisely constrain cosmology. Collett \& Auger \citep{collett2014} modeled the HST imaging, performing a pixelated reconstruction of both sources, and made a {1.1\%} measurement of $\beta$. Converting this into constraints on dark energy, this single 
DSL, with a cosmic microwave background (CMB) 
data prior from \citep{planckxvi}, constrains a constant dark energy equation of 
state to $\sigma(w)\approx0.2$. 
This number is from Collett \& Auger \citep{collett2014} and actual data. While we do not focus here on parameter estimation, the important point is that 
future surveys are forecast to increase the known galaxy scale population by $\sim100\times$  \citep{Collett2015}, with a similar increase in the compound lens population including DSL.

The Euclid satellite \cite{euclid} should discover $\sim1700$ galaxy-scale DSL systems that are suitable for cosmology. 
Rubin Observatory's Legacy Survey of Space and 
Time (LSST) is forecast to discover a similar number of systems \citep{MandelbaumDESCSRD}. 
(Galaxy scale systems strongly lensing more than two sources are much rarer and we do not include them.) These forecasts are derived from the {\sc LensPop} package \citep{Collett2015} modified to include multiple background sources.

The {\sc LensPop} approach assumes that all lens galaxies are singular isothermal ellipsoids, with velocity dispersions and ellipticities drawn from the population of elliptical galaxies observed by the Sloan Digital Sky Survey \citep{0611607}. Potential lenses are assumed to be uniformly distributed in co-moving volume. Sources are elliptical exponential profiles, with number densities and colours drawn from the LSST galaxy simulations of \cite{Connolly}.

Observations of these idealized lens systems are then simulated by mocking the LSST and Euclid point spread function and background noises. A lensed source is deemed to be detected if the signal-to-noise is greater than 20, the Einstein radius is greater than twice the seeing and the magnification is greater than 3. 
When simulating DSLs with {\sc LensPop}, we neglect the mass effect of the first source \citep{1510.00242}. We also use more stringent constraints than the {\sc LensPop} defaults: we insist that both sources have one or more image arcs of length $0.3$ arcseconds -- this ensures a reasonable possibility that the density slope of the lens can be recovered from high-resolution imaging alone. 

Since follow-up is needed for cosmological inference, we restrict ourselves to the smaller sample of 87 DSL forecasts, representing the number estimated to be discovered in the best-seeing single epoch imaging of LSST. These DSL were used in the LSST Science Requirements Document \citep{MandelbaumDESCSRD} and represent a conservative lower limit for the number of DSL that will have the  high-resolution imaging and spectroscopic follow-up that is critical to do precision cosmology with LSST DSLs. 
(The James Webb Space Telescope and ground based extremely large telescopes could further substantially improve the constraining power of each system.) We overplot the simulated systems in 
Fig.~\ref{fig:isocontours}, for those within 
the axis range, with the radius of the circles 
proportional to $1/\sigma(\beta)$, so more 
visible data are more precise. 
The choice of 87 DSLs is likely to be extremely conservative by the late 2020s: Euclid is expected to find 1700 DSLs, and will provide high resolution imaging of each of them. The 4MOST Strong Lens Spectroscopic Legacy Survey will provide spectroscopic redshifts for tens of thousands of lenses, including compound lenses. Together these surveys will deliver a much larger sample of DSLs without the need for additional followup.

Cosmological forecasts also require us to know the precision with which $\beta$ will be constrained in each DSL. Simulating this inference is beyond the scope of this work, we instead adopt a simple approximation:  we assume that $\beta$ in each system can be independently constrained with a fractional precision of the quadrature sum of the uncertainties 
$0.01({\rm arcsec}/\theta_{E,1})$, $0.01({\rm arcsec}/\theta_{E,2})$, and 0.01. The first two terms mimic the uncertainty with which Einstein radii are likely to be measured with Euclid whilst the extra 
1\% sets a floor from the uncertainty on the inferred density profile of the lenses. The results with HST \cite{collett2014} and simulations of Euclid time delay lenses \cite{meng15} indicate that these uncertainties are not unreasonable. In fact, 
they may be somewhat conservative -- compare 
the 1.1\% precision obtained in 2014 \cite{collett2014}.

\section{Constraining Dark Energy} \label{sec:DEconstraints} 

\begin{figure}[h]
\centering 
\includegraphics[width=0.45\columnwidth]{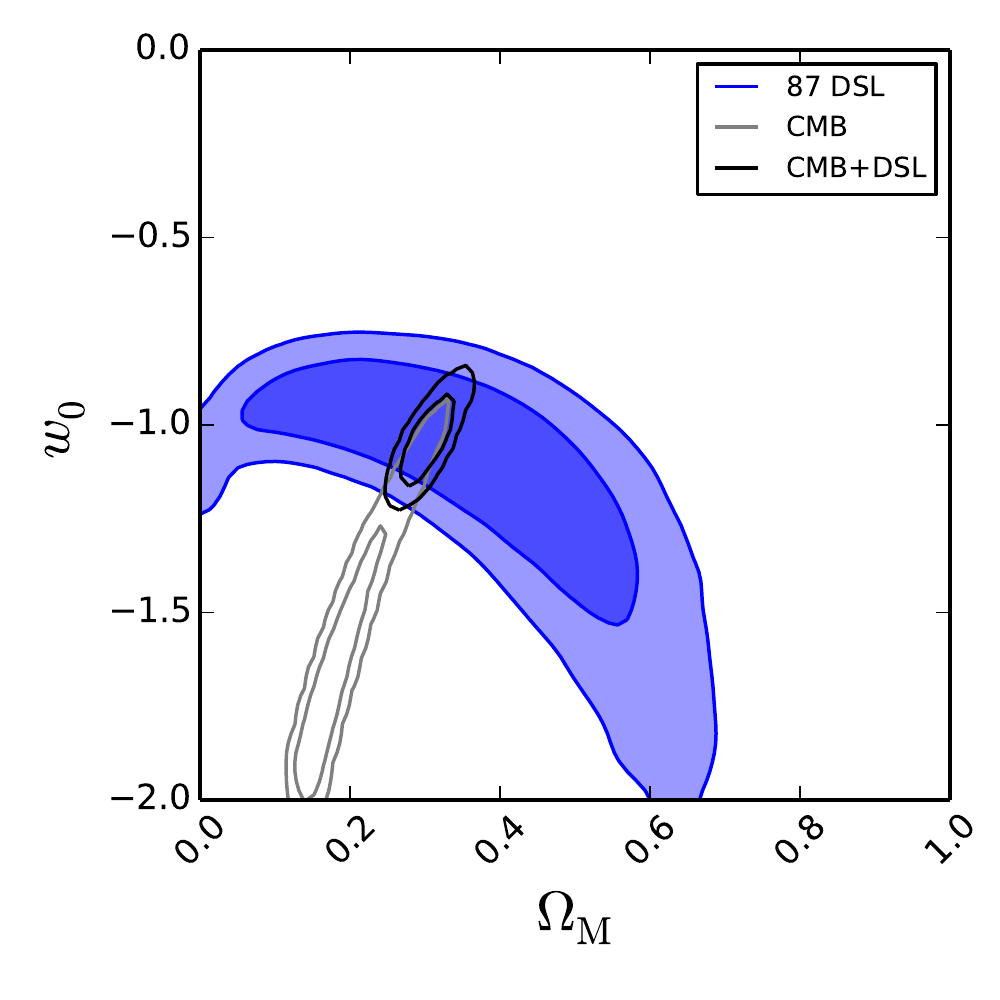}
\includegraphics[width=0.45\columnwidth]{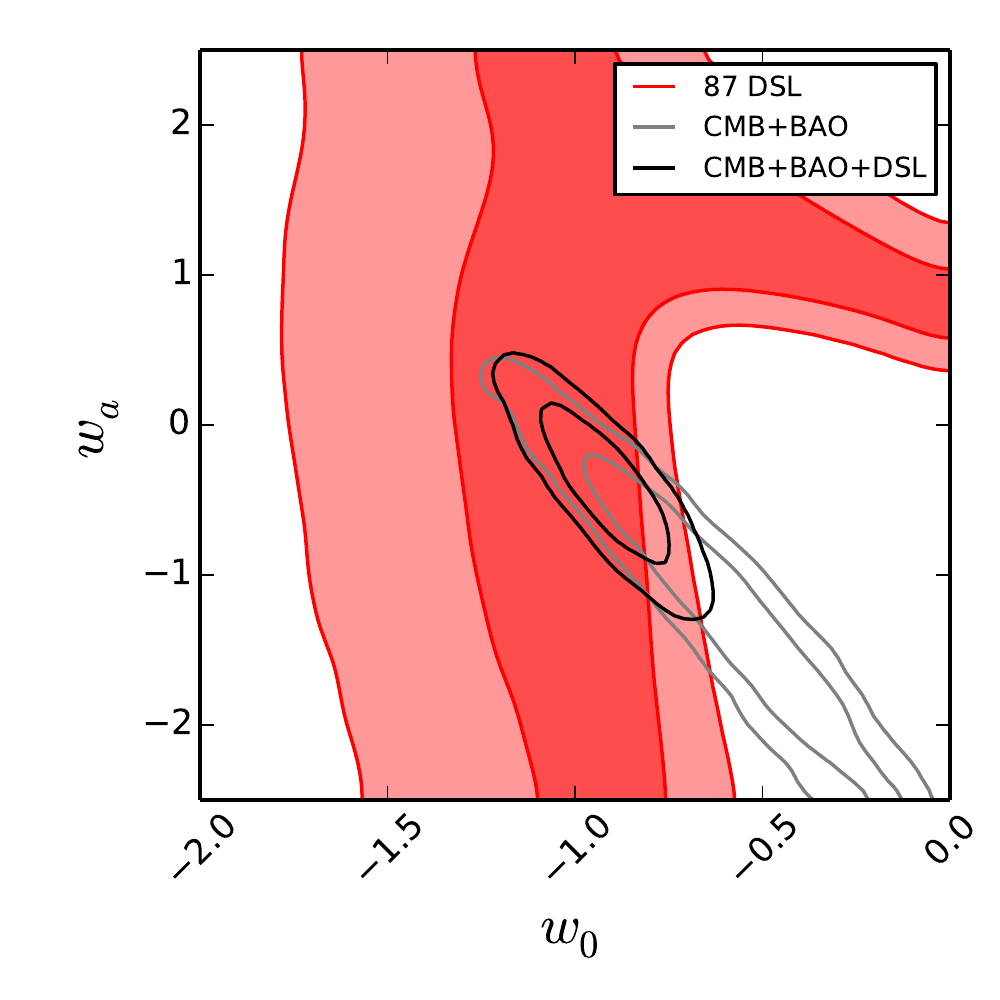}
\caption{68\% and 95\% confidence constraints on cosmological parameters. Red shows results from the 87 DSLs considered in this work, grey shows Planck 2018 
\citep{planck2018} 
(plus BAO for $w_0 w_a$CDM) and black shows the combination of both. Left panel shows the $\Omega_M-w$ plane assuming a flat $w$CDM cosmology. 
Note the DSL probe alone can give a clear indication of an accelerating universe. 
Right panel shows the $w_0-w_a$ plane, marginalizing over $\Omega_M$ in a flat $w_0 w_a$CDM cosmology. 
The upper right branch will be ruled out 
due to violation of matter domination at high redshift.}
\label{fig:mcmc} 
\end{figure}

As was pointed out in \cite{collett2012} and \cite{1605.04910}, DSL provide an independent and complementary cosmological 
probe from the standard ones. While they will be 
highly valuable in the consistency tests of the 
following sections, and as complementary data 
and as crosschecks, we briefly leave our model independent focus and explore their direct 
cosmological parameter leverage due to their 
unique combination of distance ratios. 
Figure~\ref{fig:mcmc} presents 
the results of Markov chain Monte Carlo constraints 
from 87 simulated DSL as described in 
Sec.~\ref{sec:forecast} on the matter 
density and dark energy equation of state. 
For a spatially flat $w$CDM cosmology (with a constant dark energy equation of state, $w$, not fixed to $-1$) and assuming a uniform prior on $\Omega_M$ between 0 and 1, we find $w$ = $-1.09^{+0.15}_{-0.29}$. 
Allowing the equation of state to vary with the scale factor of the universe yields a constraint on its current value of $w_0$ = $-1.01^{+0.31}_{-0.32}$, whereas the derivative is poorly constrained: $w_a$ =  $0.6^{+1.0}_{-2.0}$ (assuming a uniform prior between -2.5 and 2.5). Much of the posterior weight is in regions of the parameter space that are already strongly excluded by other cosmological data sets. This is particularly the case for the region with both $w_a \gtrsim 1$ and $w_0 \gtrsim -0.7$ which is ruled out by evidence for the existence of a matter dominated era in our Universe. For regions of the parameter space close to  $\Lambda$CDM, an interesting behavior is observed: the $w_0$ constraint is almost insensitive to $w_a$. Assuming $|w_a|<0.5$ yields very similar constraints on $w_0$ as setting $w_a \equiv 0$. 
For parameter fitting (as opposed to our 
main focus of model independent tests of 
the cosmological framework), much of the 
leverage of DSL will be in breaking degeneracies 
from other probes.

\section{Spatial Curvature Consistency Test} \label{sec:curv}

In addition to the standard cosmology parameter 
constraint estimation, we can also explore 
consistency tests of the FLRW framework. In this  
section and the following one, this is our 
main focus. 

Our universe appears to be consistent with being 
spatially flat, i.e.\ zero spatial curvature $k$ in 
the Robertson-Walker metric
\be 
ds^2=-dt^2+a^2(t)\left[\frac{dx^2}{1-kx^2}+x^2(d\theta^2+\sin^2\theta\,d\phi^2)\right]\ , 
\ee 
where $a(t)$ is the scale or expansion factor and 
$x$ is the coordinate distance while the first quantity 
in the spatial part of the metric gives the comoving 
distance element (squared) while the second quantity 
gives the usual spherically symmetric angular part. 
We can also write $k$ in 
terms of a spatial curvature quantity 
$\Ok=-k/(a_0^2H_0^2)$, where $H_0$ is the Hubble scale today 
(and we are free to define $a_0=1$). Since $k$ 
has the dimensions of 1/(distance)$^2$ then one can 
write an effective curvature energy density 
$\Omega_{\rm curv}(z)=\Ok a^{-2}=1-\Omega_{\rm total}(z)$ 
scaling as 
$a^{-2}\sim (1+z)^2$, where $z$ is the redshift.  
Here $\Omega_{\rm total}$ is the total energy density, the 
sum of the component energy densities. 

Two important questions then are whether $k=0=\Ok$, 
hence $\Omega_{\rm total}(z)=1$ for all times, 
and whether the Robertson-Walker metric lying at 
the heart of the FLRW framework is correct in taking 
$k$, equivalently $\Ok$, as constant. One can certainly attempt to 
determine the value of the $\Ok$ parameter from 
observational data and there is a large and manifold 
effort to do so (e.g.\ \cite{2205.10869,2008.11286} and many others). Instead we 
focus on the framework itself: if we determine $\Ok$ 
from the data through a relation 
between observables (rather than parameter fitting), 
is it the same value at different redshifts (and 
then secondarily we can ask if that value is zero). 
This is an example of a consistency test: taking a  
relation that must hold in FLRW and testing whether 
it is in fact always valid. 

Geometric probes, i.e.\ those that depend only on 
the Robertson-Walker metric and not on the evolution 
of matter inhomogeneities, tend to be especially 
clean testing grounds. While distances to objects 
depend on the spatial curvature, they also depend on 
the energy density of all components and these need 
to be accounted for to obtain constraints on the 
curvature from the distance data. 
The consistency tests we refer to instead involve 
relations {\it among\/} distances that hold in any FLRW model -- 
that is, one goes directly from measured distances 
to a consistency test of spatial curvature without 
adopting a specific model for the energy density contents. 

This can be thought of as triangulating points on 
a surface to determine its curvature (see discussions 
by \cite{weinberg72,1802.04816}), and so we see 
that we need not simply distances from the observer, 
but distances between remote points to form a triangle. 
Such remote distances can be provided by gravitational 
lensing, where the distance between the lens and the source 
enters. Curvature consistency tests 
directly using observables 
have explored the 
use of time delay distance from strong lens systems, 
e.g.\ \cite{1412.4976,1802.04816,1802.05532,1905.09781,1910.10365}. 
This quantity does however have sensitivity to $H_0$ and the 
lens mass model. 

The dimensionless ratio of distance ratios $\beta$ 
from double source lensing is an interesting alternative 
(it was briefly considered in \cite{1905.09781} but 
used to fit the value of $\Ok$ rather than for a redshift 
dependent consistency test). Since $\beta$ involves 
four distances -- the distance to each of the two sources 
from the observer, $r_{s1}$ and $r_{s2}$, and from the 
common lens, $r_{ls1}$ and $r_{ls2}$ -- this 
method actually uses observations to form a concave 
quadrilateral rather than a triangle. 

The remote comoving distances are related to the 
curvature in the FLRW framework by 
\be 
r_{ls}=r_s\sqrt{1+\Ok r_l^2}-r_l\sqrt{1+\Ok r_s^2}\ . 
\ee 
In terms of the observables $\beta$ and the 
various distances we can solve for 
the spatial curvature $\Ok$. While we obtain 
$\beta$ directly from the measurement of a DSL system, 
and can get distance measurements from standardized 
candles (Type Ia supernovae: SN) or rulers (baryon 
acoustic oscillations: BAO), those distances will not 
in general be 
at the precise redshifts of the lens and sources in 
the DSL system. We, therefore, need a nonparametric 
method of obtaining the desired distances; frequently 
Gaussian processes are employed for this 
(see, e.g., 
\cite{1009.5443,1204.2272,1710.04236,2206.15081} 
and many others). 

Given the observational data, the spatial curvature 
consistency relation for all redshifts is 
\be 
\Ok = \frac{(1 - \beta)^4 r_l^{-4} - 2(1 - \beta)^2 r_l^{-2}\left(\beta^2 r_{s2}^{-2}+r_{s1}^{-2}\right) + \left(\beta^2 r_{s2}^{-2}-r_{s1}^{-2}\right)^2}{4\beta(1 - \beta)\left[\beta r_{s2}^{-2}+(1 - \beta) r_l^{-2}-r_{s1}^{-2}\right]} \ . \label{eq:omegakfull}
\ee 
Again we emphasize that this must hold in the 
FLRW framework, independent of the specific energy 
density components and parameter values. 
Any statistically significant deviation from constancy 
of $\Ok$ points to either a measurement systematic or 
a violation of the FLRW framework. And a constant 
value of $\Ok\ne0$ obtained over a wide range of redshifts would 
point to potentially more robust evidence against flatness 
than a standard parameter fit analysis. 

From measurements of the distances and $\beta$  
we can propagate their uncertainties to $\Ok$. 
Since they come from 
different measurements, plus the redshifts are well separated, 
it is a reasonable assumption to take the uncertainties to be  
independent. Thus we can add them in quadrature when 
propagating them to $\sigma(\Ok)$. 

\begin{figure}[h]
\centering 
\includegraphics[width=0.6\columnwidth]{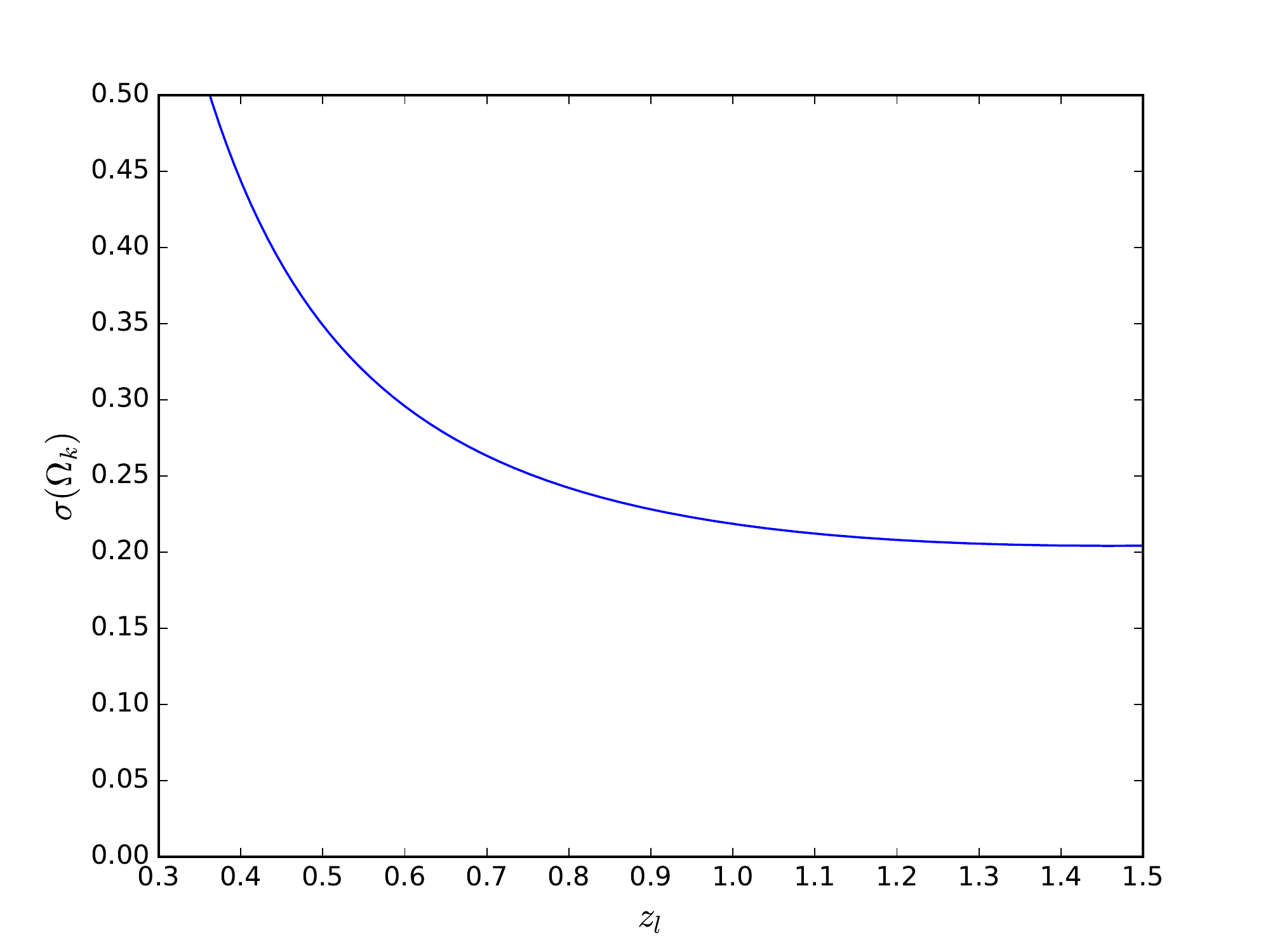}
\caption{The uncertainty $\sigok$ from a single DSL 
with $\beta$ measured to 1\% (plus comoving distance 
measurements to 1\% from SN and BAO data) is plotted 
vs lens redshift $z_l$. The uncertainty will roughly 
square root down with the number of observations at each 
redshift, down to some systematics limit. }
\label{fig:sigmaok} 
\end{figure}

Figure~\ref{fig:sigmaok} shows the result for 1\% precision  
on each of the four observable quantities, with $\beta$ 
taken to be from a single DSL system having the $z_l$ shown. 
For simplicity we set $z_{s1}=2z_l$ and $z_{s2}=1.5z_{s1}$ as reasonable rules of thumb (this puts $\beta$ in the broad 
valley of Fig.~\ref{fig:isocontours}, and see also \cite{2204.03020}). 
Note that $\sigok$ improves for higher 
redshift systems, leveling off at $\sigok\approx0.2$ for $z_l\gtrsim0.9$. 
The uncertainty will reduce 
with multiple DSL, so (if all uncertainties can be reduced statistically) for example 16 DSL at $z_l=0.9$ will 
deliver $\sigok\approx0.05$ if the measurements are independent, and this can be carried out 
at multiple lens redshifts. 
The actual result will depend on where the 
systematics limits exist for the 
distances and the DSL distance ratio. 
High redshift lenses at $z_l\gtrsim0.8$, such 
as should be readily found by Euclid, will 
be especially useful.

Following \cite{1802.04816} we can define a 
redshift dependent curvature quantity $K$ that is 
zero for a flat universe. 
This ``$K$ test'' gives a somewhat different window on
spatial curvature in that $K$ has a specific redshift dependence
predicted within the FLRW framework. Therefore one can use any measured
deviations from this -- as a function of redshift -- as 
an 
indicator of issues with the framework, or observations. 
The  $K$ test in terms of the observables is 
(compare Eq.~\ref{eq:betadef}) 
\bea  
K &\equiv& \beta - \frac{1 - \frac{r_l}{r_{s1}}}{1 - \frac{r_l}{r_{s2}}} \\ 
&\approx& \frac{r_l}{2}\frac{r_{s2}}{r_{s1}}\left(\frac{r_{s1} - r_l}{r_{s2} - r_l}\right)(r_{s2} - r_{s1})\Omega_k + \mathcal{O}(\Omega_k^{2})\ . \notag 
\eea  
Note that we always use the full form of $K$ to test curvature, 
with the first-order expansion just shown for intuition. 
In addition to the different redshift weighting
mentioned above, the K test also has a different weighting on data 
than Eq.~\eqref{eq:omegakfull} 
and so it has the potential 
to provide a crosscheck or different view of curvature. 

\begin{figure}[h]
\centering 
\includegraphics[width=0.49\textwidth]{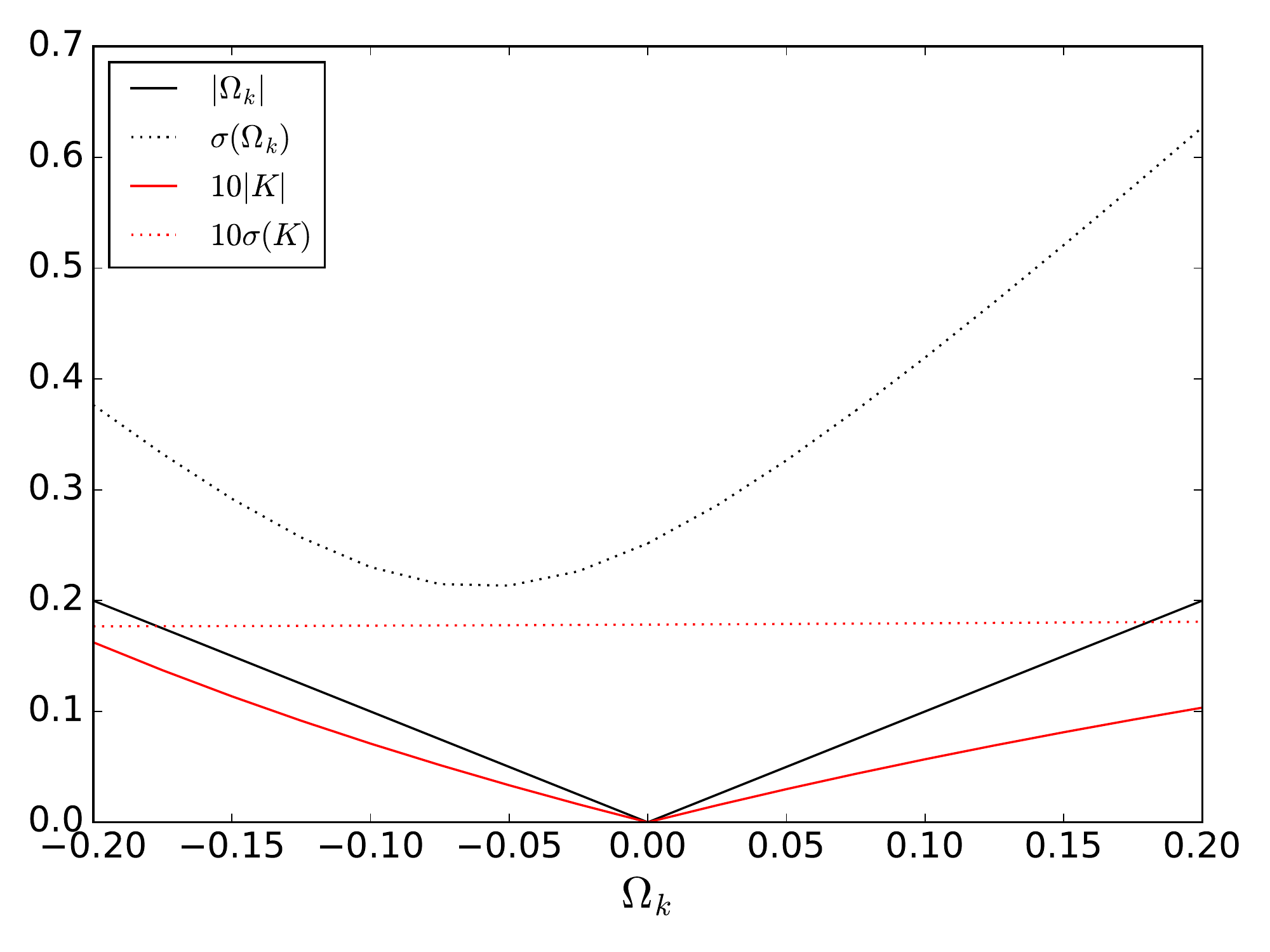}\ 
\includegraphics[width=0.49\textwidth]{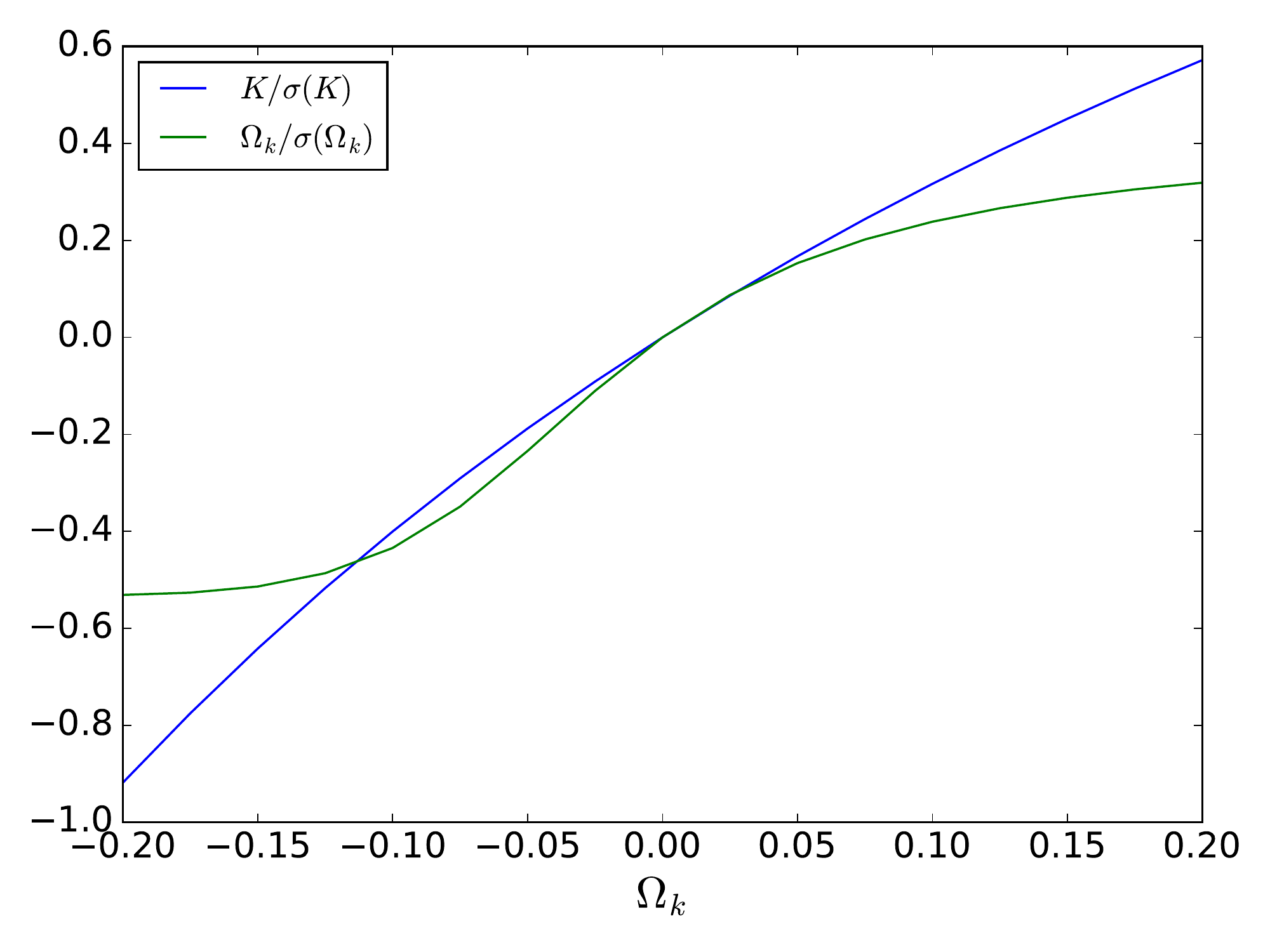}
\caption{[Left panel] Absolute values and uncertainties of the two curvature quantities $\Omega_k$ and $K$, derived from measurements of 
$\beta$, $r_l$, $r_{s1}$, $r_{s2}$, plotted vs the input $\Omega_k$. 
The uncertainties (dotted curves) are for a 
single DSL system at $z_l = 0.75$ , $z_{s1} = 1.5$, $z_{s2} = 2.25$ with fractional distance uncertainties on $\beta$, $r_l$, $r_{s1}$ and $r_{s2}$ of 1\% each. 
[Right panel] The effective signal-to-noise S/N of each curvature test (with the sign of the signal intact) is plotted vs $\Ok$. 
The two curvature tests have nearly the same S/N over the range $\Ok=[-0.1,+0.05]$. The uncertainty should roughly decline as $1/\sqrt{N_{\rm sys}}$ for $N_{\rm sys}$ systems measured, and thus the S/N increase; e.g.\ $\sim25$ systems measured should give signal-to-noise of 1 on determining $\Ok=-0.05$, for example.}
\label{fig:fig1} 
\end{figure}

Figure \ref{fig:fig1} shows these two curvature quantities and their uncertainties as a function of $\Omega_k$ for a DSL system 
with $z_l = 0.75$, $z_{s1} = 1.5$, $z_{s2} = 2.25$. 
As before, for simplicity, we adopt a fractional $\beta$ precision of 1\%, and lens and source distance precisions (from, e.g., supernova or BAO distances) of 1\%. The intersection point between the curvature quantity ($\Omega_k$ or $K$) curve and its uncertainty curve determines the lower bound on the 
ability to measure the curvature parameter for a given, single 
DSL system. This is made more explicit in the 
right panel showing signal to noise like 
ratios. While a single DSL system has $S/N<1$, 
multiple DSL can be used in these curvature 
consistency tests. Note that unlike the 
time delay distance consistency test of 
\cite{1802.04816}, the quantities $\Ok$ and 
$K$ do carry somewhat distinct information, 
with $K$ having more leverage at higher 
curvature amplitudes. One can certainly 
imagine applying one curvature consistency 
test with time delay distances and performing a crosscheck 
on any violation with DSL distance ratios.

While statistically the uncertainty estimation on 
curvature will never be as good as a model dependent 
parameter estimation (Planck \citep{planck2018} provides 0.016 uncertainty 
for $\Lambda$CDM, but this loosens by a factor two 
for $w$CDM, and even more so for $w_0w_a$CDM), these consistency tests do 
provide an alternate, model independent approach. 
Such a more robust method will be highly valuable 
if evidence is presented from parameter estimation 
of either deviation from flatness or especially 
violation of the FLRW framework.

\section{Distance Duality Consistency Test} \label{sec:dual} 

The DSL distances ratio involves angular diameter distances 
entering the light deflection. We can explore a consistency 
test between angular diameter distances and luminosity 
distances, such as from Type Ia supernovae, 
in a similar way to what we did in the previous section with 
spatial curvature. Converting both to comoving distances, we 
have 
\be 
r_i^{SN}=d_L/(1+z_i) \quad; \quad r_i^\beta=d_A\,(1+z_i)\ . 
\ee 
We expect consistency in terms of what is often called the distance 
duality relation $d_L=d_A\,(1+z)^2$, 
attributable to Tolman, Ruse, 
Etherington 
\cite{tolman,ruse,etherington} in various degrees 
of generality and with formal 
proofs by \cite{ellis,weinberg72}). This relation should 
hold when four very general conditions are valid 
(also see the pedagogical discussion in \cite{2105.02903}): 
1) metricity, 2) geodesic completeness, 3) photons propagate 
on null geodesics, and 4) adiabaticity. Violation of any one 
would have revolutionary consequences for cosmology. 

Nevertheless, we can test this by writing 
\be 
r_i^{SN}=r_i^\beta\,(1+z)^\eps \ , 
\ee 
and seeing if $\eps=0$ is consistent with data. 
(See, e.g., \cite{2210.04228,2010.04155,1906.09588,1511.01318} for a selection of 
other analyses using strong lensing.) 
That is, from supernova distances we predict what 
$\beta$ we should get, and compare it to the observed 
$\beta$. However, we emphasize that any discrepancy is 
not automatically interpretable as a violation of 
distance duality. We therefore consider and compare 
three sources of any such discrepancy: 1) violation of 
the distance duality, $\eps\ne0$, 2) incorrect cosmological model assumed, 
specifically $\Ok\ne0$, and 3) systematics. 

Let's begin with systematics. As an 
example we consider astrophysics in the form of 
an evolution in 
power law slope of the lens mass profile with redshift, 
\be 
\gamma=2+\mu\left[\left(\frac{1+z_l}{1.4}\right)^2-1\right]\ . 
\ee 
Thus, the mass density profile near the Einstein radius, 
$\rho\sim r^{-\gamma}$, may have $\gamma=2$ (the singular 
isothermal sphere, or SIS, value) at $z=0.4$, but 
evolves from approximately $2-\mu/2$ at $z=0$ to 
$2+2\mu$ at $z=1.5$. This is just a toy model to 
illustrate systematics, e.g.\ for $\mu=0.05$ the 
slope would evolve from $\gamma\approx 1.975$ to 2.1. 
As the mass density profile affects the light deflection 
angle and hence Einstein radius, measuring $\beta$ 
as the ratio of Einstein radii would give 
\be 
\beta(\mu)=\left[\beta(\mu=0)\right]^{1/[\gamma(\mu,z_l)-1]} \ . 
\ee 

For considering the effect due to spatial curvature, to evaluate $\beta(\Ok)$ we simply 
use the appropriate 
distances for a universe with curvature parameter $\Ok$, 
as in the previous section. 

Finally, in the case of violation of distance duality, 
the distances ratio predicted by supernova distances is 
(for a flat universe with $\gamma=2$ lens mass profile 
-- we vary each potential source 
of discrepancy one at a time) 
\be
\label{eq:betaeps}
\beta(\eps) = \frac{1 - \frac{r_l}{r_{s1}}\left(\frac{1+z_l}{1+z_{s1}}\right)^\epsilon}{1 - \frac{r_l}{r_{s2}}\left(\frac{1+z_l}{1+z_{s2}}\right)^\epsilon} \ \ . 
\ee 

\begin{figure}[h]
\centering 
\includegraphics[width=0.6\columnwidth]{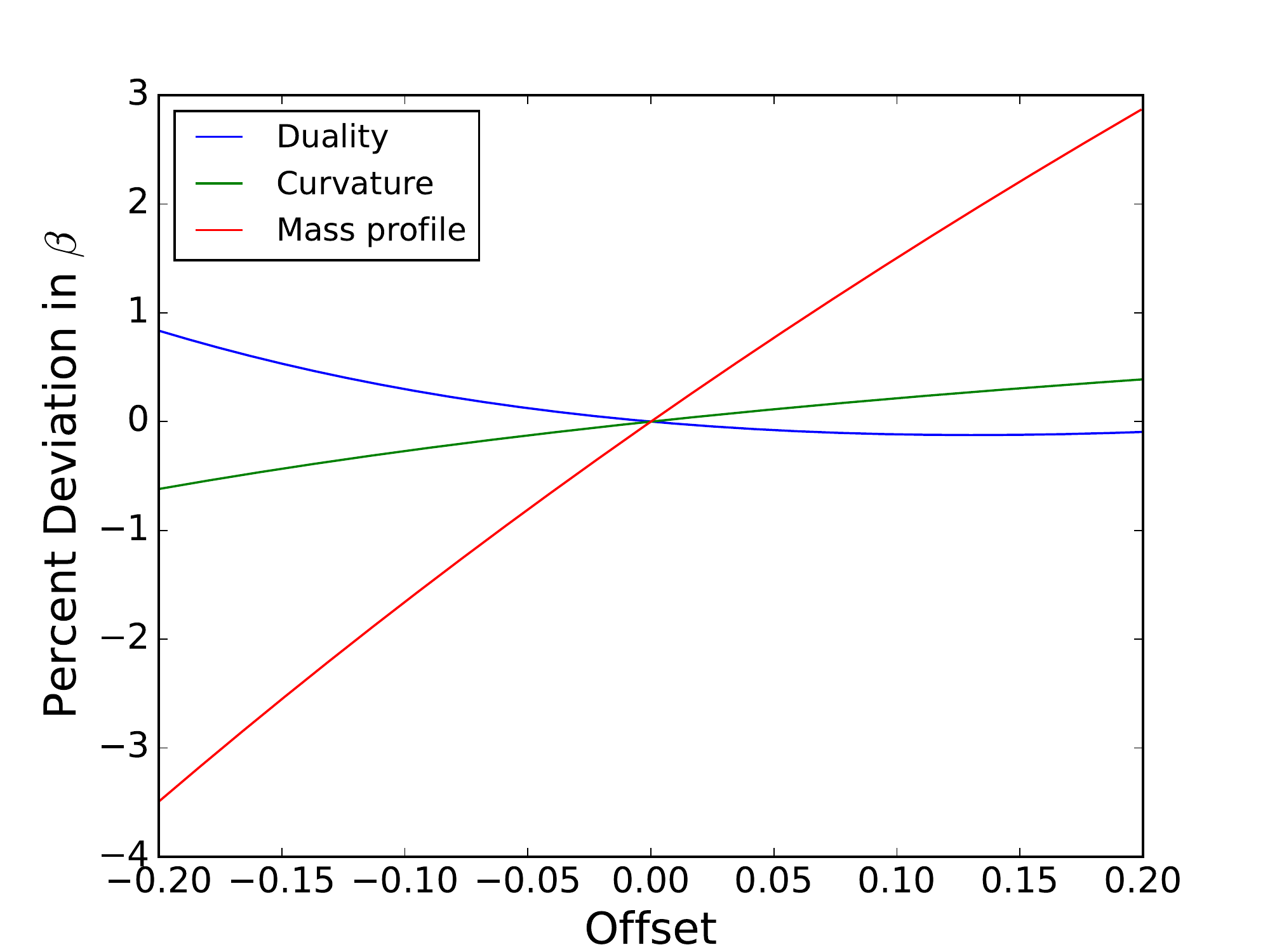}
\caption{Percent deviation in $\beta$ is plotted as a function of offset of each parameter one at a time 
from fiducial values zero, for distance duality 
violation parameter $\epsilon$, spatial curvature $\Omega_k$, 
and mass profile evolution $\mu$, 
for a DSL system with $z_l=0.75$, $z_{s1}=2z_l$, $z_{s2}=3z_l$. 
} 
\label{fig:distduality} 
\end{figure}

We can then compare the fractional deviation 
\be 
\left(\frac{\delta\beta}{\beta}\right)_p=\frac{\beta(p)-\beta(0)}{\beta(0)}\ , 
\ee 
for each of $p=\eps$, $\Ok$, $\mu$. 
Figure~\ref{fig:distduality} shows the percent deviation 
in $\beta$ as a function of each offset, for the particular 
case of a DSL system with $z_l=0.75$, $z_{s1}=1.5$, 
$z_{s2}=2.25$. 
The imparted deviations due to distance duality violation 
and curvature are mostly comparable in 
magnitude, but are swamped by the necessity of tight 
systematics control. That is, 
before making any claim for violation of distance 
duality, one must ensure that the knowledge of the 
cosmological model (here specifically spatial 
curvature) and systematics is respectively at least as 
or much more rigorously known. 

\begin{figure}[h]
\centering 
\includegraphics[width=0.6\columnwidth]{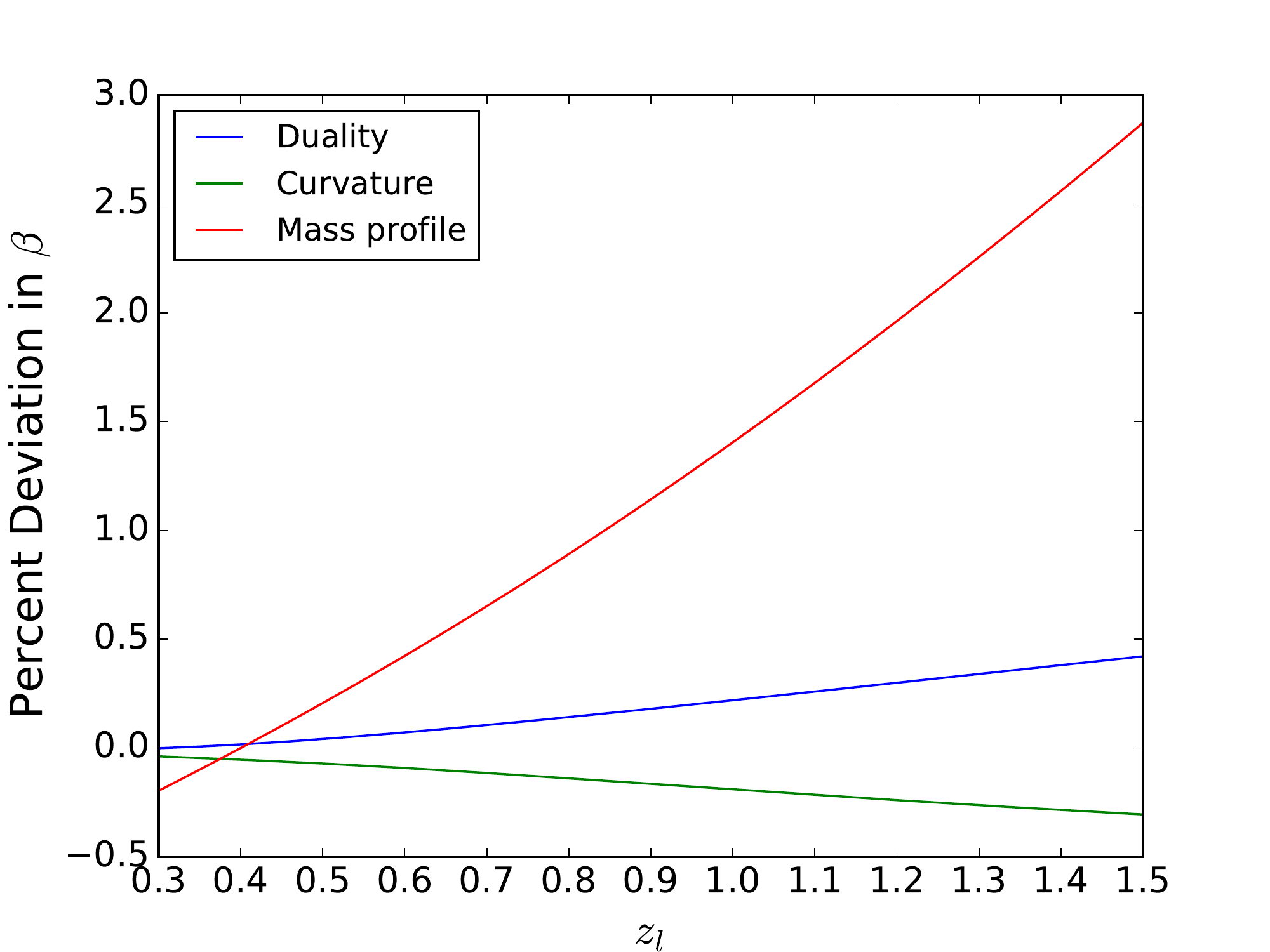}
\caption{Percent deviation in $\beta$ as a function of lens redshift (keeping $z_{s1}=2z_l$, $z_{s2}=3z_l$) is shown for offsets from zero 
of magnitude $-0.05$, $-0.05$, $+0.05$ respectively 
for distance duality violation $\eps$, spatial curvature $\Ok$, 
and mass density profile evolution $\mu$. 
} 
\label{fig:distdualityVszl} 
\end{figure}

Figure~\ref{fig:distdualityVszl} shows the dependence 
on $z_l$ (keeping the same ratios $z_{s1}/z_l=2$ 
and $z_{s2}/z_{s1}=1.5$) of the percent deviation, 
for the case of $\eps=-0.05$ or $\Ok=-0.05$ or $\mu=+0.05$. 
The impacts are roughly linear in $z_l$. Again we find 
that equal amplitudes of offset in 
distance duality violation or curvature 
cause nearly the same 
deviation in $\beta$ regardless of which parameter is 
offset; however, systematics must be particularly tightly 
controlled, by roughly five times better. 
Thus  the cosmological model and 
especially astrophysical and experiment systematics 
have to be well-known before new physics 
such as a distance duality violation can be pursued.

One could constrain the lens mass profile slope, and 
hence $\mu$, with follow-up observations of the lenses, 
e.g.\ through spectroscopy to measure lens kinematics. 
This is certainly the best approach, but we end by exploring 
the internal ``self calibration'' of the lens mass profile through 
measurements of $\beta$ itself. For this, we restrict to 
a standard, flat cosmology and seek to fit $\mu$, together 
with the cosmological parameters $\Om$, $w_0$, and $w_a$, 
using the mock sample of 87 LSST-like DSL systems. For 
a quick estimate we use the information matrix formalism, 
assume uniform 1\% uncertainty on each $\beta$, and study 
the effect of external priors from the CMB or from 
cosmic surveys in the form of an effective prior on $\Om$ of 0.01. 

We find the lens mass profile is not strongly covariant with 
the cosmology parameters, with correlation 
coefficients $r_{\mu i}<0.84$ for each cosmology 
parameter $i$, in all cases. Marginalized constraints 
are $\sigma(\mu)\approx0.022$ for no or either prior, 
improving to 0.015 if both priors are applied. Note 
that $\mu=0.022$ corresponds to the mass profile slope 
evolving over $\gamma=[1.99,2.05]$ from $z=0$ to 1.5. 
From Fig.~\ref{fig:distduality}, however, we see that 
this would still swamp a comparable amplitude  
violation of distance duality. Restricting to $\Lambda$CDM 
we find $\sigma(\mu)\approx0.012$ for no or all priors. 

\begin{figure}[!bth]
\centering 
\includegraphics[width=0.6\columnwidth]{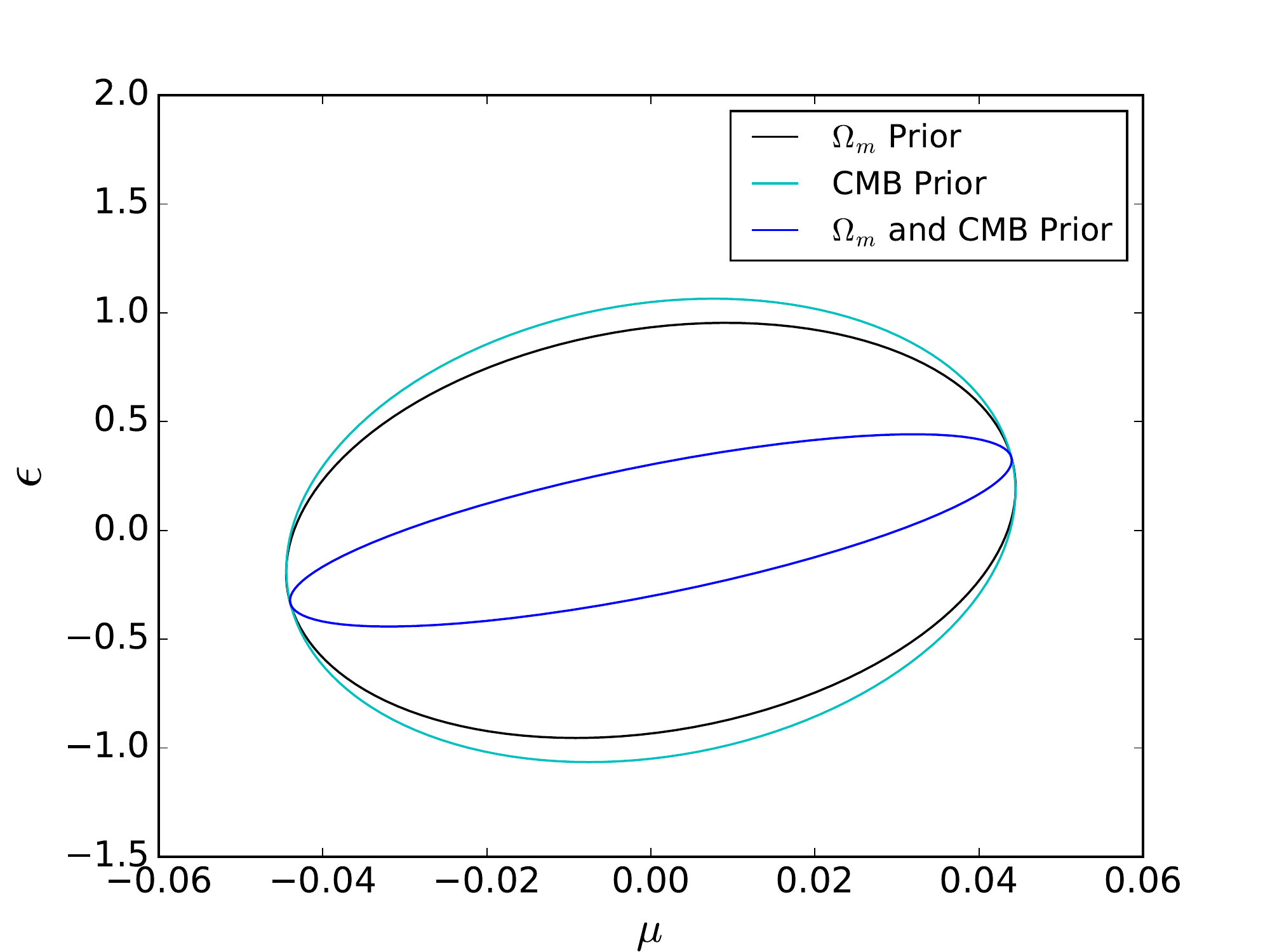}
\caption{1$\sigma$ joint confidence contours on the 
distance duality violation parameter $\eps$ and the lens mass profile evolution parameter $\mu$ expected from the 87 DSL considered in this work, with various priors. Note the significant covariance between the parameters, i.e.\ evidence of new physics is only possible with stringent systematics control.} 
\label{fig:MuEps} 
\end{figure}

If we do want to carry out the distance duality test robustly, 
simultaneously fitting for the lens mass profile evolution and (flat) cosmology, we indeed find strong 
covariance of the distance duality parameter $\eps$ with the mass profile evolution $\mu$. 
Figure~\ref{fig:MuEps} shows the degeneracy between $\mu$ and $\epsilon$ using the 87 DSL forecasts. 
One can readily see that a vertical cut at $\mu=0$ (i.e.\ fixing to no evolution) gives much tighter 
constraints on $\eps$ than in the case marginalizing over $\mu$. 
The 1$\sigma$ marginalized uncertainties are $\sigma(\epsilon) = 0.628$, $\sigma(\mu) = 0.029$ 
for the case with an added $\Omega_m$ prior of 0.01. Adding a CMB 
prior gives similar results, with 
$\sigma(\epsilon) = 0.701$, $\sigma(\mu) = 0.029$, while 
using both priors delivers $\sigma(\epsilon) = 0.291$, $\sigma(\mu) = 0.029$.

\section{Conclusions} \label{sec:concl} 

Double source lensing offers a promising probe of 
cosmology, including the FLRW framework itself. It has the 
useful characteristics of ``remote viewing'' the 
universe with some distances 
independent of the observer, of being geometric, 
substantially independent of structure growth, and 
of being dimensionless, independent of the Hubble constant. 

The number of DSL will increase some 1000 fold with 
the advent of wide-field surveys such as Euclid and 
LSST, though the number of the most robust systems 
will be limited by follow-up resources, particularly with the inevitable demise of the Hubble Space Telescope, which has provided much of the high resolution imaging for strong lensing cosmology so far.
Even so, we find that a data set of fewer than 100 DSL can have 
significant impact on not only cosmology parameter 
estimation (an independent constraint 
on the dark energy equation of state $w$ = $-1.09^{+0.15}_{-0.29}$, with this result essentially insensitive to the variation of the equation of state with cosmological scale factor) 
but also, the focus of this article, testing  the fundamental cosmological 
framework. Furthermore, DSL has good complementarity 
with other probes \cite{2204.03020}, including time 
delay distances from other strong lensing systems 
\cite{1605.04910}. 

The DSL property of measuring the distance between 
distant objects allows a form of triangulation 
(really quadrangulation) enabling tests of the 
homogeneous, isotropic FLRW framework. One can map 
spatial curvature quantities as a function of lens 
redshift -- in a model-independent manner, especially 
valuable since parameter estimation approaches to 
$\Omega_k$ tend to be sensitive to what other parameters 
are allowed to vary. Moreover we show consistency 
relations that investigate whether {\it each\/} 
system gives $\Omega_k=0$ -- 
testing flatness -- and whether they are all constant -- testing FLRW. 
While a small number of systems gives modest constraints, 
DSL provides an independent consistency test from other 
probes (such as time delay distances), and one can 
build up larger samples given follow-up resources. Moreover, DSL serves as an 
important crosscheck if a violation of FLRW is detected 
by another probe, essential for such a revolutionary 
result. 

We show how to check the consistency of foundational 
light propagation properties, such as that light 
travels on null geodesics, through the distance 
duality test. However, we caution, and 
quantitatively demonstrate, that other cosmological and 
astrophysical effects can mimic the violation of the distance 
duality relation if not properly accounted for, such 
as spatial curvature and evolution in the lens mass 
profile slope. The latter dominates in general, and 
such a systematic would need to be controlled at the 
$\sim1\%$ level in order to cleanly observe a distance 
duality violation even as large as $\sim5\%$. 

With upcoming surveys delivering new discoveries 
of DSL and other lens configurations, further 
probes may be developed as well. With DSL's 
ratio of distance ratios, and reduced systematics from  
lens modeling, the field of strong gravitational 
lensing beyond the standard one source--one lens paradigm 
looks quite promising, and worthy of further attention.

\acknowledgments 

DS acknowledges the support provided by the Berkeley Physics Undergraduate Research Scholarship (BPURS) from the Department of Physics, University of California, Berkeley.  TEC. is funded by a Royal Society University Research Fellowship and the European Research Council (ERC) under the European Union's Horizon2020 research and innovation program(LensEra: grant agreement No. 945536). EL is supported by the Energetic Cosmos Laboratory, 
by NASA ROSES grant 12-EUCLID12-0004, 
and by the U.S.\ Department of Energy, Office of Science, Office of High Energy Physics, under contract no.\ DE-AC02-05CH11231.  For the purpose of open access, the authors have applied a Creative Commons Attribution (CC BY) licence to any Author Accepted Manuscript version arising. Supporting research data are available on reasonable request from the corresponding author.


\end{document}